**Bias and response heterogeneity in an air quality data set**

S. Stanley Young, CGStat, genetree@bellsouth.net
Robert L. Obenchain, Risk Benefit Statistics LLC, wizbob@att.net
Christophe Lambert, Center for Global Health, Division of Translational Informatics, Department of Internal Medicine, University of New Mexico, Albuquerque, New Mexico, USA

## Abstract

It is well-known that claims coming from observational studies often fail to replicate when rigorously re-tested. The technical problems include multiple testing, multiple modeling and bias. Any or all of these problems can give rise to claims that will fail to replicate. There is a need for statistical methods that are easily applied, are easy to understand, and are likely to give reliable results. In particular, simple ways for reducing the influence of bias are essential. In this paper, the Local Control method developed by Robert Obenchain is explicated using a small air quality/longevity data set first analyzed in the New England Journal of Medicine. The benefits of our paper are twofold. First, we describe a reliable strategy for analysis of observational data. Second and importantly, the global claim that longevity increases with improvements in air quality made in the NEJM paper needs to be modified. There is subgroup heterogeneity in the effect of air quality on longevity (one size does not fit all), and this heterogeneity is largely explained by factors other than air quality.

## Introduction

There is extensive literature on the question: Does air quality have health effects? For example, Google Scholar gives 199k hits for ("mortality" and "air pollution"). See **Health Effects Institute (2010)** and an editorial by **Brauer and Mancini (2014),** for example. One paper that appeared in 1993 has over 6000 citations. The vast majority of these papers find a positive association between air quality and health effects (death). A few papers make the case that if potential bias is carefully taken into account then there is no association between air quality and deaths, e.g. **Chay et al. (2003), Enstrom (2005), Janes et al. (2007), Greven et al. (2011), Cox et al. (2013)**. Clearly the weight of evidence is for a positive association, but for any particular type of claim, logically it takes only one true negative to negate all the positive associations with respect to causation for that claim. A real, causative claim should be detectable in well-designed and properly run experiments. What were some of the factors that lead to discordant findings in air quality literature?

First, there is the empirical observation that medical observational studies most often fail to replicate when claims are re-tested in Randomized Clinical Trials, **Young and Karr (2011)**. Part of the problem is the way statistical methods are typically applied to observational data, **Young and Miller (2014), Gelman and Loken (2014)**. Papers cited by these authors note very high rates of failure to replicate. There appear to be two statistical problems: first, usually multiple questions are addressed with no adjustment for multiple testing and second, multiple (usually linear models) are employed to "best fit" the data without taking into account model selection bias. Multiple modeling was pointed to as a serious problem by **Clyde (2000)**. In this paper, we focus on bias as the pervasive problem yielding claims that fail to replicate.

A relatively new statistical strategy, Local Control (LC), **Obenchain and Young (2013)**, is used here to reanalyze an air quality data set obtained from C. Arden Pope, III, Pope et al. (2009). There it is essentially claimed that observed decreases in $PM_{2.5}$ over 20 years (roughly ~1980 to ~2000) at 211 locations across the US uniformly lead to increases in longevity. The first step in LC analysis is to cluster the locations based on demographic and socio-economic covariates. Importantly, this initial clustering ignores all data on both the outcome of interest, in our case increase in longevity, and the exposure of interest, the improvement in air quality resulting from a decrease in $PM_{2.5}$, a measure of the amount of small particles in the air. The basic idea of Local Control analysis is to cluster experimental units, in our case cities, into subgroups that have similar socio-economic characteristics but are not necessarily spatially contiguous. In other words, clustering forms groups of cities that can be thought of as "statistical clones". We then make statistical comparisons primarily within these clusters of cities that are relatively homogeneous in terms of important confounding factors.

In the case of a binary choice between two treatments, LC could calculate a treatment difference t-statistic within each cluster. In the case of a continuous exposure variable, LC could use linear regression within clusters. It has been shown that Local Control analysis will better control for covariate bias in observational medical studies, Obenchain (2009, 2010) and here we apply the method to environmental epidemiology studies. Cities within a cluster have been relatively well matched with respect to measured covariates. In addition Local Control does not assume any specific model form for the adjustments for covariates. An additional claimed advantage is that the analysis process is more simple and objective ...arguably less subjective than traditional model fitting.

**Pope data set**

We summarize the nature of the Pope data set, **Pope et al. (2009)**. The dataset consists of 211 cities/regions in the US. The differences in 11 variables were computed from ~1980 to ~2000. See Table 1 taken from **Young and Xia (2013)**. The outcome variable is increase in longevity, LE. The predictor of interest is change in $PM_{2.5}$. N.B.: Pope gives a decrease in $PM_{2.5}$ as a positive value. Covariates include COPD and lung cancer as surrogates for smoking, income, education, and a number of other variables. **Table 2** from **Young and Xia (2013)** gives the variable names, means and standard deviations of the variables. We follow the Pope convention that a decrease in $PM_{2.5}$ takes positive sign. See **Figure 1,** a redrawing of his Figure 4. LE_i is an increase in longevity. PM_d is a decrease in $PM_{2.5}$. **Krstic G. (2012)**, and **Young and Xia (2013)** have re-analyzed the Pope data set. We found an outlier in the dataset, (RowID #128, 5600_NYOR_NY) and that record is excluded from our analysis here. The data set we use for this paper is deposited at datadryad.org. **[Link to be provided.]**

**Local Control Regression analysis method**

The essential features of the Local Control, LC, approach, **Obenchain (2009, 2010)** for analysis of large, observational data sets are easy to explain, even to nontechnical audiences. The original LC assumes that health care outcomes for two alternative treatments for the same disease or condition are to be compared head-to-head. LC starts by dividing all patients, without regard for their status as either treated or control, into many subgroups. Local Control Regression, LCR, also clusters the data into subgroups. Any method that assures that, within each subgroup, patients are relatively well matched on their observed baseline x-characteristics can be used; we use hierarchical clustering. A simple linear regression, SLR, is computed within each cluster where the range of the predictor variable is large enough and the sample size is suitable. The resulting SLR intercepts and slopes are subjected to further analysis. Interest is usually on the slopes as they measure the effect of the exposure variable on the response. Slopes come with test statistics. A histogram depicting the empirical distribution of t-statics is informative as are the corresponding multiple raw and adjusted p-values.

**Local Control analysis of Pope data set**

What follows is an application of Local Control Regression to the Pope data set. LCR starts by clustering the data set based on a set of covariates. The response variable and the predictor variable are not used in the clustering. A step-wise regression was used to suggest variables for use in the clustering. Change Income, Lcan_d, copd_d and black_d were highly significant predictors and were used for the clustering. The dataset was clustered into 10 clusters. **Table 3** gives the regression coefficients, t-tests and p-values for the regressions for each of the ten clusters. Intercept and slope were computed within each cluster. With the exception of cluster 4.  the local regressions of LE on $PM_{2.5}$ were not significant. While the fitted slopes for clusters 6 and 7 were larger, numerically, the slope for cluster 4 had the smallest standard error because its cluster was the largest (39 locations.) **Figure 2** gives a histogram of local slope estimates for the 10 clusters. This figure illustrates that the empirical distribution of local slope estimates spans a range that includes negative as well as positive values.

The variability of the $PM_{2.5}$ slope estimates from Local Control Regression displayed in **Table 3** and **Figure 2** suggests that these effect-sizes could come from a traditional probability distribution. However, to establish that these are truly **heterogeneous treatment effects**, it is essential to demonstrate that they are so-called "fixed" (expected, predictable) effects rather than purely random effects (unpredictable variation.) To establish heterogeneity, we will show that the full range of local slope estimates are predictable from the socio-economic characteristics of the individual locations.

Predictions of observed effects typically result from fitting a multivariable regression model, and the "goodness" of the resulting fit is typically judged (subjectively) from the scatter plot of the estimates versus their predicted values. Unfortunately, such regressions can make strong (highly restrictive) but wrong assumptions about smoothness of continuous functional relationships.  This  typically results in poor fits (low R-squares) that fail to convince skeptics that effects truly are predictable (heterogeneous.)  Thus, here, we will focus on a regression "tree" fitting method, recursive partitioning (RP), that is highly flexible and easily interpretable as well as a statistically valid approach to prediction. Much improved fits (high R-squares) of RP predictions from rather small trees to LC local slope and intercept estimates will result.

To perform these RP analyses, the estimated intercepts and slopes for each cluster are appended to the observations in that cluster. That gives a 210 observation data set. We first used SAS JMP to perform Ward clustering on the standardized values of socio-economic confounding characteristics of the 210 locations to form 10 subgroups. For a description of the Recursive Partitioning methods used here see **Hawkins (2009)**. Briefly, all the data are examined to find a single "best" confounder and the cut point" that will divide the data into two "more homogeneous" subgroups. The SAS JMP implementation allows only two-way splitting. Each subgroup is examined in turn, and the data set is recursively split until some stopping criterion is met. Several layers of splitting are given in **Figures 3&4** for intercepts and slopes respectively. The larger the LogWorth the more likely the split is not due to randomness. We consider a split significant if the LogWorth is greater than 3 (p-value of 0.001). The Intercept RP analysis has an $R^2$ of 0.828 while the Slope RP tree has a $R^2$ of 0.647; these are exceptionally good fits for such small trees. Since confounders can predict the simple linear regression intercepts and slopes, the heterogeneity of air pollution associations is strongly supported. In short, the intercepts and slopes of the within cluster regressions are well predicted from the socio-economic covariates given in the Pope data set.

N.B.: There are no published papers where Local Control or Local Control Regression analysis has been applied to environmental epidemiology data sets.  LC has been extensively validated on simulated and actual medical observational data sets; See **Obenchain (2009, 2010)**, **Obenchain and Young (2013)**, **Lopiano et al. (2014), Faries et al. (2013)**. Recursive partitioning is widely used for analysis of complex data.

**Discussion**

Local Control Regression, LCR, is a new method based on the Local Control research by Dr. Obenchain. The essence of the Local Control, LC, approach, **Obenchain and Young (2013)**, is to cluster the data into subgroups with similar confounder characteristics and then compute local (conditional) measures of treatment response within each cluster.  In our case, changes in longevity potentially due to changes in $PM_{2.5}$ exposure are locally estimated using simple linear regression. The resulting intercepts and slopes are then predicted across clusters using information about environmental confounding factors, e.g. income, COPD, etc. LC assures that, within each cluster (subgroup), observations are relatively well-matched on confounding covariates. The quality of this matching can be assessed.

Conceptually, LCR is very similar to what **Janes et al. (2007)** and **Greven et al. (2011)** have done; they look within and across locations, and we work within and across clusters of locations. Our intercepts represent baseline effects while our  slopes can be taken to be the adjusted effects of longevity versus $PM_{2.5}$ exposure within clusters. The LCR analysis we propose is also in the spirit of **Pope et al. (2009)**. The data that C. Arden Pope, III, has shared with us are only the changes in longevity, covariates and exposure from ~1980 to ~2000; see their Figure 4 on NEJM page 382. The ratio of increase in LE to decrease in $PM_{2.5}$ is thus the slope of the line connecting two time points in the original data for each of 210 locations. The NEJM publication did not discuss the global intercept in Figure 4. One might consider re-analyzing the full Pope data set (see Figures 2 and 3, NEJM pages 380-381) so that heterogeneous intercepts and intermediate values can be examined.

We say again, "one size does not fit all." In nine of ten clusters there is no *prima facia* evidence for an effect of the reduction of $PM_{2.5}$ leading to an increase in longevity. If there is to be any regulation of $PM_{2.5}$, then regulation should be limited to locations where $PM_{2.5}$ reduction actually appears to unambiguously increase longevity. The four left-most clusters depicted in our Figure 2 contain 59 locations where, on average, changes in longevity tended to be negative relative to observed decreases in $PM_{2.5}$. For some of these 59 locations, undue emphasis on improvements in air quality may have left insufficient local funding for changes in socio-economic covariates that would have had a more positive impact on longevity.

LC strategy focuses attention upon local effect-size estimates and the full range of their observed empirical distribution, rather than upon the statistical significance of some of their values. LC estimates almost always vary in numerical size and usually also vary in precision due to differences in cluster sizes dictated by observed confounder distributions as well as by differences in observed exposure variation in LCR (or in local treatment choice fractions in traditional LC.) In fact, the only phase of LC where p-values play potentially important roles is in the final, parametric model selection phase where local effect-sizes are being predicted from observed confounding characteristics of experimental units.

Having said that one size fits all makes no sense, we turn to an examination of Cluster 4 and the strength of the evidence for a significant association of longevity to $PM_{2.5}$ decrease. The raw p-value of 0.0007 and the adjusted p-value of 0.0071 are impressive by usual standards. However, there are recent arguments that to help ensure reproducible results, p-values need to be much smaller, **Boos and Stefanski (2011), Johnson (2013)**. Both make the case that a p-value addressing a single question should be 0.005 or even 0.001 or even smaller for a claim to be considered reliable. The adjusted p-value of 0.0071 misses so there is an argument for chance.

However, the fact that recursive partitioning analysis of the slopes and intercepts using the confounding variables was successful argues that these intercepts and slopes are not uniform across covariates; their differences are not random, and effects of confounders on longevity are the main issue in the data.

So, where do we stand? The small p-value for Cluster 4 could be random, due to known confounders or due to unmeasured confounders. The Local Control Regression followed by Recursive Partitioning points to confounders rather than $PM_{2.5}$. From 1957 to 1966 on the Perry Mason TV show, lawyer Perry Mason would not only get his client off of a murder charge, but also he would also identify the actual murderer. We have three suspects. Since both the intercepts and slopes computed in the clusters are highly predicted from covariates, $R^2$s of 0.828 and 0.647 respectively, the evidence strongly points to confounders. While we have cast serious doubt on the Pope et al. claim that lowering of $PM_{2.5}$ increases longevity uniformly across the entire US, it remains for future research to identify the true culprit in Cluster 4, if any. We can say that however the data is examined, lowering of $PM_{2.5}$ is not causative of a uniform increase in longevity.

The data set used in this research can be obtained at datadryad.org. Supplemental material can also be obtained at arxiv.org. **[Specific URLs will be provided later.]**

**References**

Boos, DD, Stefanski LA. (2011), "P-value precision and reproducibility," *The American Statistician*, 65, 213-221.

Brauer M, Mancini GBJ. (2014) Where there's smoke . . . Poor air quality is an important contributor to cardiovascular risk. *BMJ* 348:g40 doi:10.1136/bmj.g40.

Chay K, Dobkin C, Greenstone M. (2003) The Clean Air Act of 1970 and adult mortality. *J Risk Uncertainty* 27,279-300.

Clyde M. (2000) Model uncertainty and health effect studies for particulate matter. *Environmetrics* 11, 745–763.

Enstrom JE. (2005) Fine particulate air pollution and total mortality among elderly Californians, 1973–2002. *Inhalation Toxicology* 17, 803–816.

Faries DE, Chen Y, Lipkovich I, Zagar A, Liu X, Obenchain RL. (2013) Local control for identifying subgroups of interest in observational research: persistence of treatment for major depressive disorder. *Int J Methods Psychiatr Res* 22:185–194.

Gelman A, Loken E. (2014) The statistical crisis in science. *The American Scientist* 102, 460-465.

Greven S, Dominici F, Zeger S. (2011) An approach to the estimation of chronic air pollution effects using spatio-temporal information. *J Amer Stat Assoc*. 106, 396-406.

Hawkins DM. (2009) Recursive partitioning. *Comput Stat* 1, 290–295.

Health Effects Institute (2010) Proceedings of an HEI Workshop on Further Research to Assess the Health Impacts of Actions Taken to Improve Air Quality. Communication 15.

Janes H, Dominici F, Zeger S. (2007) Trends in air pollution and mortality: an approach to the assessment of unmeasured confounding. *Epidemiology*. 18, 416-423.

Johnson VE. (2013) Revised standards for statistical evidence. *PNAS* 110(48):19313–19317.

Krstic G. (2012) A reanalysis of fine particulate matter air pollution versus life expectancy in the United States, *J Air Waste Manage Assoc* 62, 989–991.

Lopiano KK, Obenchain RL, Young SS. (2014) Fair treatment comparisons in observational research. *Statistical Analysis and Data Mining* 7, 376–384.

Peng RD, Dominici F, Zeger SL. (2006) Reproducible epidemiologic research. *American Journal of Epidemiology* 163, 783-789.

Obenchain RL. (2009) SAS macros for local control (phases one and two). Observational Medical Outcomes Partnership (OMOP), Foundation for the National Institutes of Health (Apache 2.0 License). download - http://support.sas.com/publishing/bbu/zip/61876.zip (subfolder: SAS code and datasets / Chapter07_Local_Control / LC SAS macros and datasets)

Obenchain RL. (2010) The local control approach using JMP. In *Analysis of observational health care data using SAS*, ed. D. E. Faries, A. C. Leon, J. M. Haro, and R. L. Obenchain, 151–192.Cary, NC, SAS Press. download - http://support.sas.com/publishing/bbu/zip/61876.zip (subfolder: SAS code and datasets / Chapter07_Local_Control / LC JMP Scripts and datasets)

Obenchain RL, Young SS. (2013) Advancing statistical thinking in health care research. *Journal of Statistical Theory and Practice* 7, 456-469.

Pope III CA, Ezzati E, Dockery DW. (2009) Fine-particulate air pollution and life expectancy in the United States, *NEJM* 360, 376–386.

Young SS, Fogel P. (2014) Air pollution and daily deaths in California. Proceedings, 2014 Discovery Summit. https://community.jmp.com/docs/DOC-6691/

Young SS, Miller HI. (2014) Are medical articles true on health, disease? Sadly, not as often as you might think. *GeneticEngineering & Biotechnology News* May 1, 34 (9).

Young SS, Karr A. Deming, data and observational studies: A process out of control and needing fixing. (2011) *Significance* September, 122–126.

Young SS, Xia JQ. (2013) Assessing geographic heterogeneity and variable importance in an air pollution data set. *Statistical Analysis and Data Mining* 6, 375-386.

**Table 1.** Variables reported and use by Pope *et al*. for regression analysis. NB: All variables are given as change from years ~1980 to ~2000.

| Variable | Comment |
|---|---|
| Life Expectancy, life-table methods | Response variable (Change LE) |
| Per capita income (in thousands of $) | Inflation adjusted to the year 2000 (Income) |
| Lung Cancer (Age standardized death rate) | Surrogate for smoking (LCan) |
| COPD (Age standardized death rate) | Surrogate for smoking. COPD denotes chronic obstructive pulmonary disease |
| High-school graduates (proportion of population) | (hs) |
| PM2.5 ($\mu g/m^3$) | Particulate matter, aerodynamic diameter $\leq 2.5$ $\mu m$ |
| Black population (proportion of population) | Self reported (black) |
| Population (in hundreds of thousands) | (pop) |
| 5-Year in-migration (proportion of population) (mig) | Five-year in-migration refers to the proportion of the population who did not reside in the county 5 years earlier. |
| Hispanic population (proportion of population) | Self reported (hisp) |
| Urban residence (proportion of population) | (urban) |

**Table 2.** Means and standard deviations for the 11 variables in the Pope data set.

| | Variable | Mean | SD |
|---|---|---|---|
| 1 | Change LE | 2.7312 | 0.9167 |
| 2 | Lcan_d | 2.3455 | 2.7726 |
| 3 | copd_d | 4.4397 | 2.4358 |
| 4 | Change Income | 8.5069 | 3.1608 |
| 5 | Change PM | 6.5477 | 2.9151 |
| 6 | hs_d | 0.1872 | 0.1453 |
| 7 | black_d | 0.0176 | 0.0565 |
| 8 | hisp_d | 0.0333 | 0.0431 |
| 9 | Pop_d | 0.9948 | 2.2599 |
| 10 | urban_d | 0.2002 | 0.1800 |
| 11 | mig_d | −0.0063 | 0.0613 |

Table 3. Statistics on intercepts and slopes for ten clusters

| | Cluster | Intercept | SE Inter | t Inter | P Val Int | Slope | SE Slope | t Slope | P val Slope | FDR P val |
|---|---|---|---|---|---|---|---|---|---|---|
| 1 | 1 | 2.3904 | 0.3625 | 6.5938 | <.0001 | 0.0506 | 0.0498 | 1.02 | 0.3246 | 0.6556 |
| 2 | 2 | 2.8103 | 0.2357 | 11.9217 | <.0001 | 0.0158 | 0.0337 | 0.47 | 0.6429 | 0.6753 |
| 3 | 3 | 2.5456 | 0.6923 | 3.6772 | 0.0079 | -0.0616 | 0.1260 | -0.49 | 0.6398 | 0.6753 |
| 4 | 4 | 1.7328 | 0.2045 | 8.4719 | <.0001 | 0.0977 | 0.0265 | 3.69 | 0.0007 | 0.0071 |
| 5 | 5 | 1.8589 | 0.6964 | 2.6693 | 0.2282 | -0.0583 | 0.0952 | -0.61 | 0.6504 | 0.6753 |
| 6 | 6 | 1.8767 | 1.2921 | 1.4525 | 0.1803 | 0.1229 | 0.2329 | 0.53 | 0.6106 | 0.6753 |
| 7 | 7 | 1.6104 | 0.4955 | 3.2501 | 0.0035 | 0.1396 | 0.0785 | 1.78 | 0.0887 | 0.4434 |
| 8 | 8 | 1.3595 | 0.3286 | 4.1366 | 0.0004 | 0.0490 | 0.0489 | 1.00 | 0.3278 | 0.6556 |
| 9 | 9 | 4.3366 | 0.2879 | 15.0656 | <.0001 | -0.0392 | 0.0331 | -1.19 | 0.2521 | 0.6556 |
| 10 | 10 | 3.5745 | 0.2329 | 15.3454 | <.0001 | -0.0136 | 0.0322 | -0.42 | 0.6753 | 0.6753 |

Figure 1. Increase in longevity, LE_i, versus decrease in PM$_{2.5}$, PM_d. As the decrease in PM2.5 increases, there is an increase in life expectancy, LE_i.

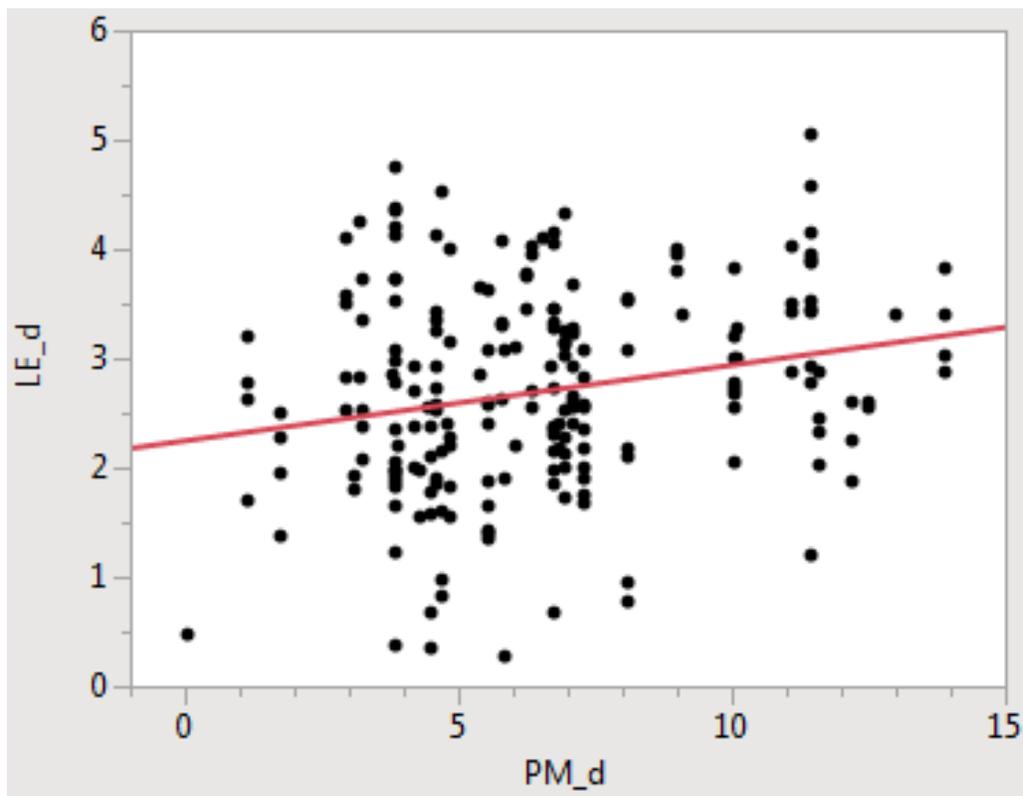

Figure 2. The Empirical Distribuion of Within-Cluster Slopes clearly spans Zero. In fact, 28% of locations (59 of 210) have negative local (adjusted) slope estimates. (Cluster Frequency Weight = Number of Locations within Cluster)

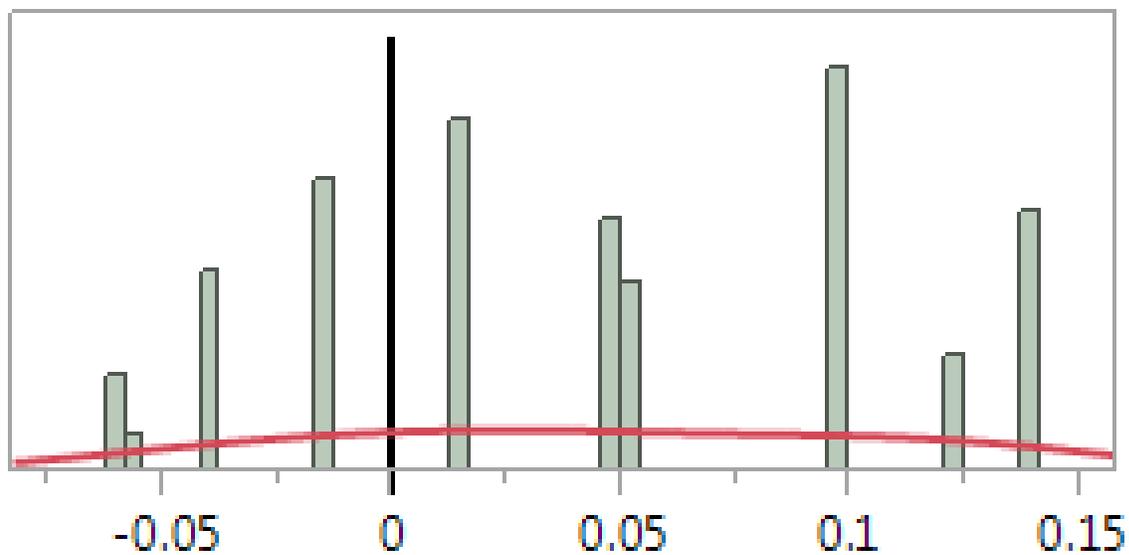

Figure 3. Fitted Recursive Partitioning tree for predicting LCR Intercept estimates.

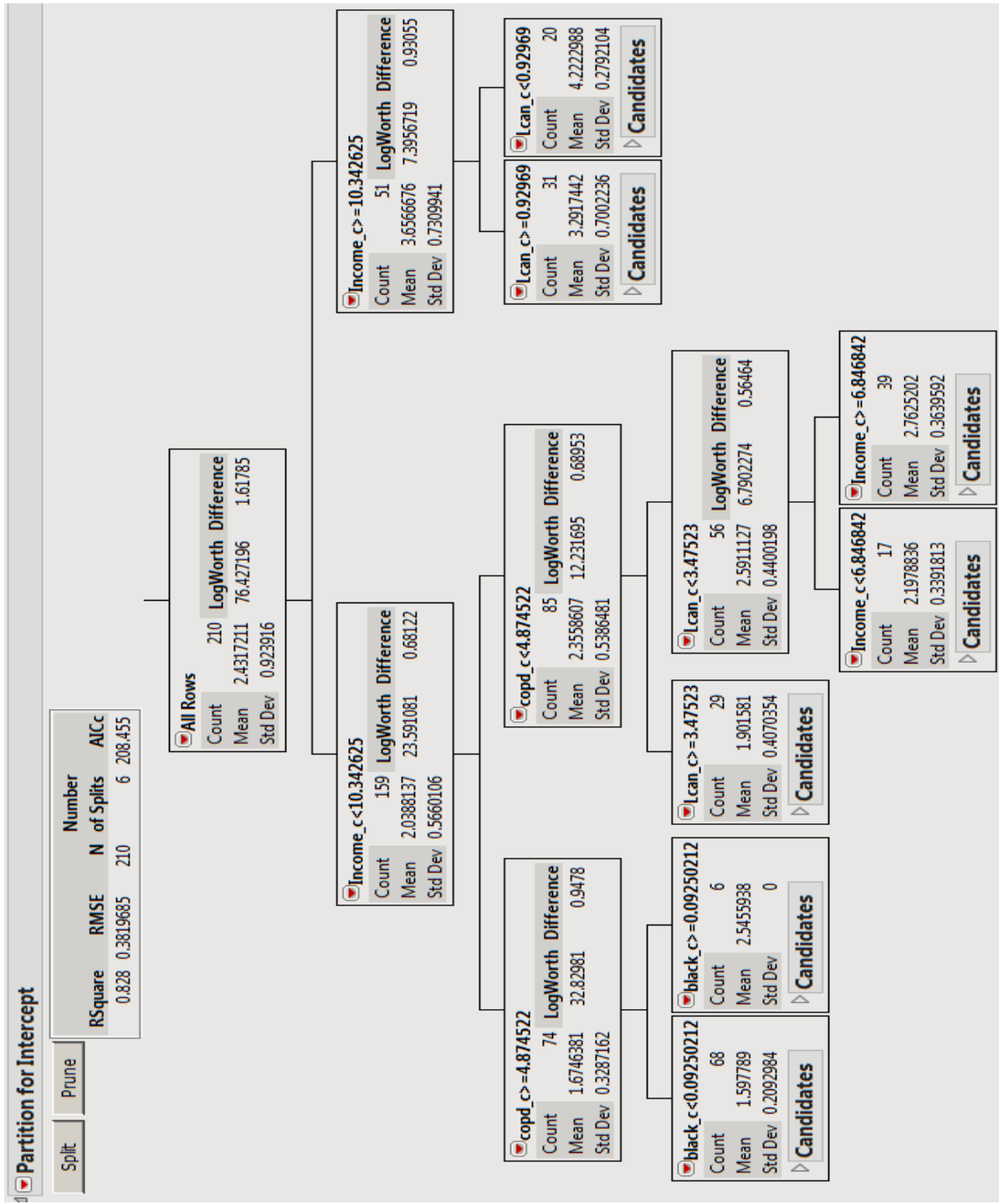

Figure 4. Fitted Recursive Partitioning tree for predicting local longevity/PM2.5 Slope estimates.

LogWorth is the negative of the log10 of the p-value for the split.